\documentclass[conference,compsocconf]{IEEEtran}

\usepackage{graphicx}
\usepackage{amssymb}
\usepackage{algorithm}
\usepackage{algorithmic}

\usepackage{eso-pic}
\AddToShipoutPicture*{\small{\sffamily\raisebox{1.8cm}{\hspace{1.8cm}{978-1-4673-5828-6/13/\$31.00\copyright2013 IEEE}}}}

\newcommand{\mynotex}[1]{}

\begin{document}

\title{Performance and Energy Efficiency of Mobile Data Offloading  with Mobility Prediction and Prefetching
}

\author{
\IEEEauthorblockN{Vasilios A. Siris and Maria Anagnostopoulou}
\IEEEauthorblockA{
Mobile Multimedia Laboratory, Department of Informatics \\
Athens University of Economics and Business, Greece\\
}
}

\maketitle

\begin{abstract}
We present a detailed evaluation of procedures that exploit mobility prediction and prefetching to enhance  offloading of traffic from mobile networks to WiFi hotspots, for both delay tolerant and delay sensitive traffic. We  consider empirical measurements and evaluate the  percentage of offloaded traffic, data transfer delay, and  energy consumption of the proposed procedures.
Our results illustrate how various factors such as  mobile, WiFi  and hotspot backhaul throughput, data size, number of hotspots, along with time and throughput estimation errors, influence the performance and energy efficiency of mobile data offloading enhanced with mobility prediction and prefetching.
\end{abstract}

\section{Introduction}

A major trend in mobile networks in the last few years is the exponential increase of powerful yet affordable personal mobile devices, such as smartphones and tablets, with multiple wireless interfaces that include 3G/4G/LTE and WiFi. The proliferation of such  devices has resulted in a skyrocketing growth of mobile traffic, which in 2011 grew 2.3-fold, more than doubling   for the fourth year in a row, and is expected to grow 18 times from 2011 until 2016\footnote{Source: Cisco Visual Networking Index: Global Mobile Data Traffic Forecast Update, 2011-2016, Feb. 12, 2012}.
On the flipside, despite its increase, the mobile data revenue significantly lags behind the exponential growth of data traffic.
One solution to address the strain from the mobile data traffic is to move a portion of it to WiFi networks, exploiting the significantly lower cost of WiFi technology and existing backhaul infrastructure.

The goal of this paper is to evaluate procedures that exploit  mobility prediction combined with WiFi and mobile throughput prediction, along with data prefetching to enhance mobile data offloading to WiFi \cite{Sir++12}. Mobility prediction can provide information on a vehicle's  route  and the time that the vehicle will reach different locations along its route. Such mobility information can be combined with geo-location information regarding WiFi hotspot access and WiFi and mobile throughput,  to predict the number of WiFi hotspots  the vehicle will encounter,  the duration of access and estimated throughput for each hotspot, and the estimated mobile throughput along the route where it will have only mobile access.
The current paper's contribution is to provide a detailed evaluation of the procedures initially proposed in \cite{Sir++12}, considering empirical measurements and investigating the impact of various factors on the overall performance. In particular,
\begin{itemize}
\item we evaluate the percentage of offloaded traffic, the data transfer delay, and the energy consumption of the proposed mobile data offloading schemes, and
\item we investigate the influence of the data object size, the mobile, WiFi, and ADSL backhaul throughput, number of WiFi hotspots, and errors in throughput and time estimation on the performance and energy efficiency.
\end{itemize}
Prior work has shown the predictability of bandwidth for cellular networks  \cite{Yao++08} and for WiFi \cite{Nic++08,Pan++09}. The work of \cite{Yao++09} investigates how bandwidth prediction can improve scheduling in vehicular multi-homed networks and
\cite{Yao++10} investigates improvements for mobile video rate adaptation.
Bandwidth prediction, together with transparent  roaming and handover, for improving video streaming is investigated in  \cite{Eve++11}.
Bandwidth prediction for client-side pre-buffering to improve video streaming is investigated in \cite{Sin++12}. The works \cite{Yao++10,Eve++11,Sin++12} focus on mobile networks, whereas our work investigates mobile data offloading to WiFi. Moreover, unlike \cite{Sin++12} which considers pre-buffering at the client device, we investigate  prefetching data to local caches in WiFi hotspots.

Exploiting  delay tolerance to increase   mobile data offloading to WiFi  is investigated in \cite{Bal++10,Ris++11}. The work of \cite{Lee++10} showed that delay tolerance of up to 100~seconds  provides minimal offloading gains; however, this applies to human daily mobility, rather than vehicles.
The work in \cite{Hou++11} applies a user utility model to reduce the mobile throughput by offloading traffic to WiFi, focusing on a transport layer protocol design to integrate cellular and WiFi networks, and utilizing throughput prediction over a 1-second  interval.
Our work differs in that we consider both delay tolerant and delay sensitive traffic, and exploit data prefetching and  prediction involving multiple WiFi hotpots along a vehicle's route.

The feasibility of using prediction together with prefetching is investigated in \cite{Des++09}, which  develops  a prefetching protocol (based on HTTP range requests), but does not propose or evaluate specific prefetching algorithms. In this paper we propose algorithms for delay tolerant and delay sensitive traffic, and evaluate their performance and robustness against time and throughput estimation errors.
Prefetching to improve the performance of video delivery  is investigated in \cite{Gol++12}, which proposes a centralized model to prefetch data in cellular femtocell networks. Prefetching algorithms to reduce the peak load of mobile networks by offloading traffic to WiFi hotspots are investigated in \cite{Mal++12}. Our work differs in that we consider prefetching for multiple WiFi hotpspots along a vehicle's route, and investigate client-side algorithms for prefetching in the case of both delay tolerant and delay sensitive traffic. In this respect, our work also differs from \cite{Li++11,Han++12} which focus on identifying the subset of nodes, along with their storage, for disseminating information using opportunistic communication.



As mentioned above, prior work has shown the predictability of bandwidth for cellular networks  \cite{Yao++08} and for WiFi \cite{Nic++08,Pan++09}.
Hence, our goal is not to develop a new system for mobility and bandwidth prediction, but to  evaluate procedures that exploit prediction information that is available by systems such as the ones mentioned above, in order to utilize prefetching  and enhance mobile data offloading to WiFi.

The rest of this paper is structured as follows:
Section~\ref{sec:procedures} discusses the mobile data offloading procedures to exploit prediction and prefetching for delay tolerant and delay sensitive traffic. Section~\ref{sec:evaluation} evaluates the procedures considering empirical measurements and investigating how various factors impact the performance and energy efficiency. Finally, Section~\ref{sec:conclusions} concludes the paper identifying future research directions.

\mynotex{ Prior/related work on:
\begin{itemize}
\item Offloading of delay tolerant traffic
\item Wifi throughput prediction
\item Mobile throughput prediction, e.g. for streaming multimedia traffic.
\item Prefetching. In space-domain versus time-domain
\item mobile-WiFi handover not the focus of the paper.
\item also how prefetching can be implemented (http get) is discussed in \cite{Des++09}.
\end{itemize}
}

\mynotex{Additional notes:
\begin{itemize}
\item In this paper we investigate prediction in the time-domain, namely a nodes location in some future time. We can also exploit mobility prediction in the space-domain, i.e. to consider multiple future network attachment point where a node can move to, and prefetch data not in one location but in multiple possible future locations.
\end{itemize}

}

\section{Enhancing mobile data offloading with mobility prediction and prefetching}
\label{sec:procedures}

Next we present the procedures that exploit mobility prediction and prefetching to enhance mobile data offloading, which were originally proposed in \cite{Sir++12}. Mobility prediction provides knowledge of how many WiFi hotspots a node (vehicle) will encounter, when they will be encountered, and for how long the node will be in each hotspot's range. In addition to the aforementioned mobility information, we assume that there is information on the estimated throughput of the WiFi hotspots and the mobile network, at different positions along the vehicle's route; for the former, the information includes both the throughput for transferring data from a remote location, e.g., through an ADSL backhaul link, and the throughput for transferring data from a local cache over a WiFi link (this estimate is used only in the case of prefetching).


Prefetching can be advantageous when the throughput of transferring data from a local cache in the WiFi hotspot is higher than the throughput from the data's original  server location. This occurs when the backhaul link connecting the hotspot to the Internet has low capacity (e.g., is an ADSL link) or when it is congested; this is likely to become more common as the  IEEE 802.11n standard becomes more widespread.

\mynotex{
\begin{itemize}
\item Common feature of both is prefetching, which involves estimating the amount of cached data and the offset within the data object from which to cache.
\item Different is that in the case of delay tolerant traffic, there is a delay threshold within which the data object needs to be transferred. Hence, we try to minimize the throughput of the mobile network. On the other hand, in the case of delay sensitive traffic we use the maximum throughput of the mobile network.
\end{itemize}
}

\subsection{Delay tolerant traffic}

For delay tolerant traffic our objective is to maximize the amount of data offloaded to WiFi, while ensuring that the whole data object is transferred within a given delay threshold.
The pseudocode for the procedure to exploit mobility prediction and prefetching is shown in Algorithm~\ref{alg:delaytolerant}.
The procedure defines the computations and actions that a mobile node takes when it exits a WiFi hotspot, hence has only mobile access (Line~\ref{line:mobile}), and when it enters a WiFi hotspot (Line~\ref{line:wifi}).
Initially, the procedure estimates the amount of data that can be transferred over  WiFi  (Line~\ref{line:datawifi}), and from this the amount of data that needs to be transferred over the mobile network (Line~\ref{line:datatimemobile}). Additionally, the procedure estimates the total time the node has WiFi access (Line~\ref{line:timewifi}) and, from this value and  the delay threshold, it estimates the duration the node has only mobile access (Line~\ref{line:datatimemobile}). From the amount of data that needs to be transferred over the mobile network and the duration of  mobile-only access, the minimum  throughput for transferring data over the mobile network can be estimated (Line~\ref{line:thrmobile}). To perform prefetching,  whenever the node exits a WiFi hotspot the procedure estimates the amount of data to be prefetched (cached) in the next WiFi hotspot (Line~\ref{line:cache}) and the corresponding offset (Line~\ref{line:offset}); this offset depends on the amount of data that will be transferred over the mobile network until the node reaches the next WiFi hotspot.

When the node enters a WiFi hotspot, it might be missing some portion of the data object up to the offset from which  data has been cached in the hotspot; this can occur if, due to a time estimation error, the node reaches the WiFi hotspot earlier than the time it had initially estimated.
In this case, the missing data needs to be transferred from the data object's original remote location (Line~\ref{line:adsl_before}). Also, again due to a time estimation error, the amount of data cached in the WiFi hotspot can be smaller than the amount the node could have transferred while it is in the   hotspot's range. In this  case, the node uses its remaining time in the  hotspot to transfer data from the data object's original location (Line~\ref{line:adsl_after}).

\newcommand{\swifi}{\mbox{{\scriptsize WiFi}}}
\newcommand{\smobile}{\mbox{{\scriptsize mobile}}}
\newcommand{\sthres}{\mbox{{\scriptsize thres}}}
\newcommand{\smin}{\mbox{{\scriptsize min}}}
\newcommand{\smax}{\mbox{{\scriptsize max}}}
\newcommand{\scache}{\mbox{{\scriptsize cache}}}
\newcommand{\twifi}{\mbox{{\tiny WiFi}}}
\newcommand{\tmobile}{\mbox{{\tiny mobile}}}
\newcommand{\tthres}{\mbox{{\tiny thres}}}
\newcommand{\tmin}{\mbox{{\tiny min}}}
\newcommand{\tmax}{\mbox{{\tiny max}}}
\newcommand{\tcache}{\mbox{{\tiny cache}}}

\newcommand{\tadslno}{\mbox{{\tiny adsl}}}
\newcommand{\tbckhl}{\mbox{{\tiny bckhl}}}

\newcommand{\tnext}{\mbox{\emph {\tiny next}}}
\newcommand{\Offset}{\mbox{\emph {Offset}}}

\renewcommand{\algorithmiccomment}[1]{/* #1 */}

\algsetup{linenosize=\small}

\begin{algorithm}
\caption{Procedure to exploit mobility prediction and prefetching for delay tolerant traffic}
\begin{algorithmic}[1]
\label{alg:delaytolerant}
{\scriptsize
\STATE \textbf{Variables:}
\STATE $D$: size of data object to be transferred
\STATE $T_{\tthres}$: maximum delay threshold for transferring data object
\STATE $N_{\twifi}$: remaining  WiFi hotspots to be encountered until $T_{\tthres}$
\STATE $D^{\tmin}_{\twifi}$: estimated minimum amount of data to be transferred in all WiFi hotspots that will be encountered
\STATE $D_{\tmobile}$: amount of data  to be transferred over mobile network
\STATE $T^{\tmin}_{\twifi, i}, T^{\tmax}_{\twifi, i}$: min, max duration node is connected  to WiFi  $i$
\STATE $T_{\tmobile}$: total duration that node is not in range of WiFi
\STATE $T_{{\mbox{\tiny next WiFi}}}$: average time until node enters range of next WiFi
\STATE $R^{\tmin}_{\twifi, i}, R^{\tmax}_{\twifi, i}$: min, max throughput of WiFi  $i$
\STATE $R_{\tmobile}$: throughput to download data over the mobile network
\STATE $D^{\tcache}_{\twifi, \tnext}$: amount of data cached  in next WiFi hotspot
\STATE $\Offset$: estimated position in data object of data transferred until node enters next WiFi hotspot
\STATE \textbf{Algorithm:}
\IF { node exits WiFi hotspot } \label{line:mobile}
\STATE  $D^{\tmin}_{\twifi}  \leftarrow \sum_{i \in N_{\twifi}} \left ( R^{\tmin}_{\twifi, i} \cdot T^{\tmin}_{\twifi, i} \right )$ \label{line:datawifi}
\STATE $T^{\tmin}_{\twifi}  \leftarrow \sum_{i \in N_{\twifi}} T^{\tmin}_{\twifi, i}$ \label{line:timewifi}
\STATE  $D_{\tmobile} \leftarrow D-D^{\tmin}_{\twifi}$ \& $T_{\tmobile} \leftarrow T_{\tthres}-T^{\tmin}_{\twifi}$ \label{line:datatimemobile}
\STATE $R_{\tmobile} \leftarrow  D_{\tmobile}/T_{\tmobile}$ \label{line:thrmobile}
\STATE $D^{\tcache}_{\twifi, \tnext} \leftarrow  R^{\tmax}_{\twifi, \tnext} \cdot T^{\tmax}_{\twifi, \tnext}$ \label{line:cache}
\STATE $\Offset \leftarrow  R_{\tmobile} \cdot T_{ {\mbox{\tiny  next WiFi}}}$ \label{line:offset}
\STATE Cache $D^{\tcache}_{\twifi, \tnext}$ data in next WiFi starting from $\Offset$
\STATE Transfer data over mobile network with throughput $R_{\tmobile}$
\ELSIF  { node enters WiFi hotspot } \label{line:wifi}
\STATE Transfer data that has not been received up to $\Offset$ from original object location \label{line:adsl_before}
\STATE Transfer data from  local cache
\STATE Use remaining time in WiFi hotspot to transfer data from original object location \label{line:adsl_after}
\ENDIF
\label{line:1end}
\\ }
\end{algorithmic}
\end{algorithm}

The procedure for exploiting mobility prediction without prefetching estimates the traffic  expected to be transferred over  WiFi, and subsequently the amount of traffic that needs to be transferred  over the mobile network and  the necessary mobile throughput. The  algorithm is presented in \cite{Sir++12}.

\mynotex{
\begin{itemize}
\item Key idea is to minimize usage mobile network while assuring that the delay constraint will be satisfied.
\item Must estimate minimum amount of data to be transferred using WiFi. Calculated mobile throughput from remaining data that needs to be transferred.
\item When switch to mobile network do two things: 1) estimate mobile throughput, 2) estimate data to cache in next wifi hotspot.
\item we assume node is connected to mobile network at all times
\item when we do not perform prefetching, the transfer throughput when node is in WiFi hotspot is lower than the throughput of WiFi due e.g. to ADSL link or because throughput for transferring from remote location is smaller
\item In case prefetching is not used, the procedure just decides the throughput to use while connected to the mobile network.
\end{itemize}

}

\subsection{Delay sensitive traffic}

Similar to delay tolerant traffic, when  the mobile node exits a WiFi hotspot it estimates the offset and the amount of data that needs to be prefetched in the next WiFi hotspot that the node will encounter.
However, unlike delay tolerant traffic, in order to minimize the transfer delay for delay sensitive traffic, the node always uses the maximum throughput that is available in the mobile network.
Moreover, note that there is no procedure for exploiting only mobility prediction (without prefetching) for delay sensitive traffic, since the maximum mobile throughput is always used. The prefetching algorithm for  delay sensitive traffic is presented in \cite{Sir++12}.

\mynotex{
\begin{itemize}
\item Different with procedure in previous subsection is that we use maximum throughput of mobile.
\item Based on mobility prediction need to decide amount of data to cache in next WiFi hotspot and offset of data to cache from
\end{itemize}
}

\section {Evaluation}
\label{sec:evaluation}

\begin{figure}[t]
\centering
\includegraphics[width=3.4in]{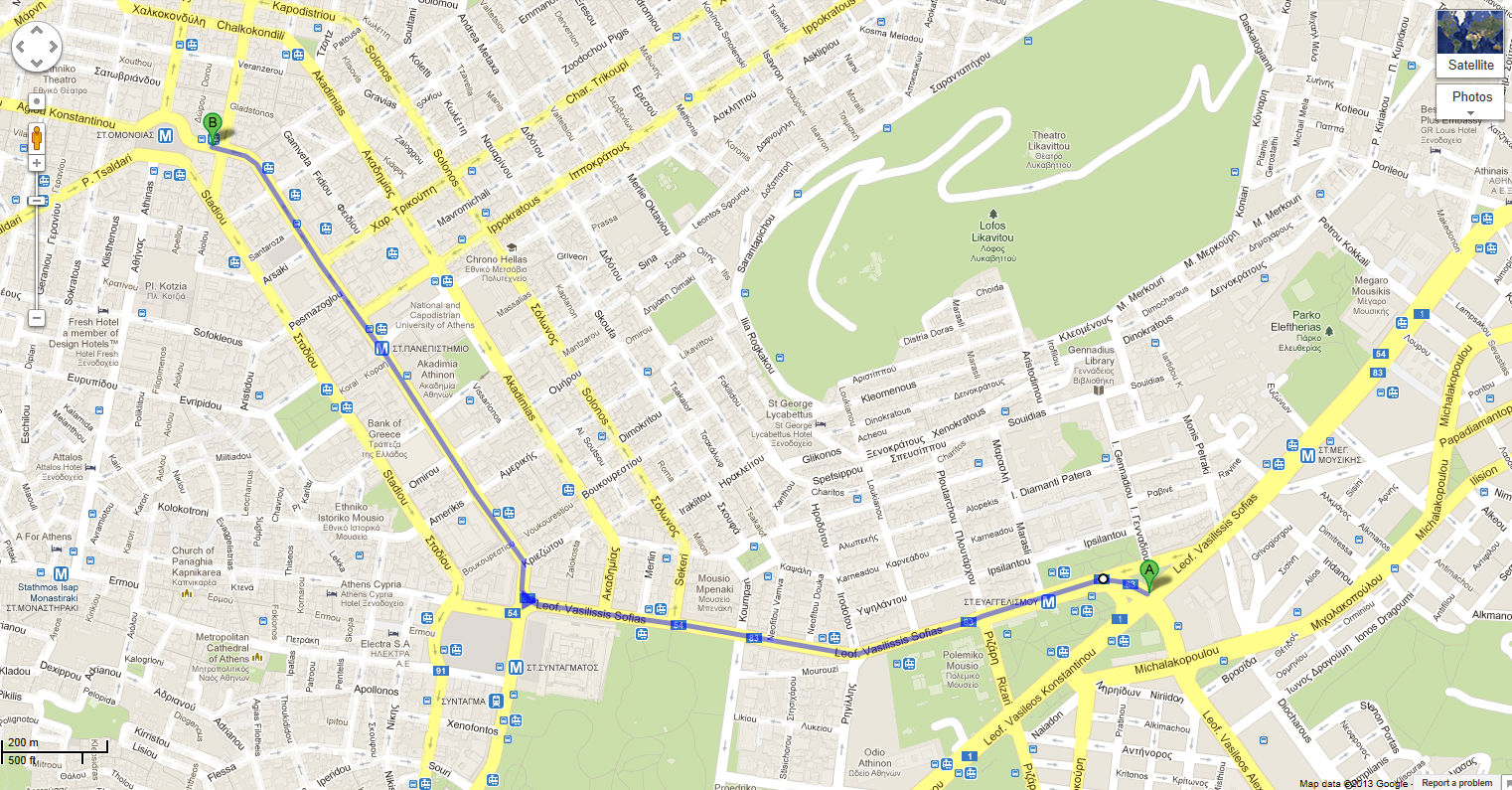}
\vspace{-0.05in}
\caption{\small{Route considered in the evaluation, along which we embed 2,4, and 8  WiFi hotspots.
The route's total travel time  is 269~seconds.}}
\label{fig:path}
\vspace{-0.15in}
\end{figure}

We consider empirical measurements for the mobile throughput and the SNR of WiFi networks along a route between two locations in the center of Athens, Greece, Figure~\ref{fig:path}, along which we
we embed 2, 4, and 8 WiFi hotspots for the various scenarios investigated.
Based on the number of hotspots  we can separate the full route into segments where the moving node has either mobile  or WiFi connectivity, as shown in Table~\ref{tab:route_segments_4} for 4 hotspots (due to space limitations, we omit the corresponding tables for 2 and 8 hotspots).

%
\begin{table}[b]
\vspace{-0.15in}
\caption{Connectivity when 4 APs are embedded along the route.  }
    \label{tab:route_segments_4}
\vspace{-0.05in}
    \centering
 {\footnotesize
\begin{tabular}{|c|c|c|c|}
         \hline
         Segment  &  Access &  Time (sec) & Throughput (Mbps)\\
         \hline \hline
     1 & mobile & 0 & 4.83\\
     2 & WiFi & 18 & 16.16 (WiFi) - 6.81 (ADSL) \\
     3 & mobile & 36 & 4.58 \\
     4 & WiFi & 90 &  16.74 (WiFi) - 8.37 (ADSL) \\
     5 & mobile & 108 & 6,1 \\
     6 & WiFi & 162 & 16.74 (WiFi) - 8.37 (ADSL) \\
     8 & mobile & 180 & 5.62\\
     9 & WiFi & 234 & 17.23 (WiFi) - 9.46 (ADSL) \\
     10 & mobile & 252 & 5.82\\
         \hline
         \end{tabular}
}
\vspace{-0.05in}
\end{table}

%

The mobile throughput  in Table~\ref{tab:route_segments_4} is the average of the values measured for each mobile segment.
However, because the WiFi APs along the route were not open, we estimated the WiFi throughput and the throughput for downloading data over an ADSL link that would have been achieved if WiFi APs were open as follows:
We initially measured the SNR value for the various APs along the route. Based on the SNR values, we estimate the throughput for downloading data stored locally at the WiFi hotspot and the throughput for downloading data over the ADSL backhaul link using Table~\ref{tab:SNR_thrpt}, whose measurements were obtained empirically from open WiFi hotspots. It is important to note that we are not suggesting that the mapping shown in Table~\ref{tab:SNR_thrpt} is universal. Rather, the above approach is used to obtain realistic throughput values that can be experienced in actual systems. Moreover, our evaluation considers different mobile, WiFi, and ADSL throughput values, as shown in Table~\ref{tab:values}, to investigate their impact on the overall performance of the mobile data offloading schemes considered.

\begin{table}[t]
\vspace{-0.05in}
\caption{Estimation of WiFi and ADSL throughput based on SNR values, using empirical measurements from open WiFi hotspots. }
    \label{tab:SNR_thrpt}
\vspace{-0.05in}
    \centering
 {\footnotesize
        \begin{tabular}{|c|c|c|}
        \hline
        SNR (dB)  &  WiFi (Mbps) &  ADSL  (Mbps)\\
        \hline \hline
    $>$ -50 & 19.90 & 15.87 \\
    -60 to -50 & 18.30 & 11.86 \\
    -70 to -60 & 17.76 & 10.13 \\
    -80 to -70 & 17.23 & 9.46 \\
    -90 to -80 & 16.74 & 8.37 \\
    $<$ -90 & 16.16 & 6.81 \\
        \hline
        \end{tabular}
}
\vspace{-0.05in}
\end{table}


The time error determines how much the times at which the node changes access technology can differ from the empirical values  in  Table~\ref{tab:route_segments_4}; for example, a 10\% time error means that the time at which the first segment (where the node has mobile access)  ends and the  second segment (where it has  WiFi access) begins is in the interval $[0.9 \cdot 18, 1.1 \cdot 18]=[16.2, 19.8]$~seconds. Note that our empirical measurements show that under typical road traffic conditions, the timing for the various route segments can differ 10-20\%.

The throughput error determines the throughput's deviation from its  average in Table~\ref{tab:values}; for example, a 40\% throughput error means that the mobile throughput is in the interval $[0.6 \cdot M,1.4 \cdot M]$~Mbps, where $M$ is the average mobile  throughput  in Table~\ref{tab:route_segments_4} which is measured empirically.
In this paper we only consider the downlink direction, hence the backhaul throughput in Table~\ref{tab:values} refers to the downstream.

%
\begin{table}[t]
\caption{Parameter values. $M, W,$ and $A$, are the mobile, WiFi, and ADSL throughput for the various segments in Table~\ref{tab:route_segments_4}}
    \label{tab:values}
\vspace{-0.1in}
    \centering
 {\footnotesize
        \begin{tabular}{|c|c|}
        \hline
        Parameter  &  Values\\
        \hline \hline
   Data object size      & 30, 40, 50 (default for delay sensitive), \\
         & 60 (default for delay tolerant), 70 MB \\
    Mobile throughput  & $M/4, M/3$ (default), $M/2, M$\\
    WiFi throughput & $W/4, W/3$ (default), $W/2, W$ \\
    Backhaul throughput & $A/4, A/3$ (default), $A/2, A$ \\
    Time error & 10\% (default), 20\%, 30\%, 40\%  \\
    Throughput error & 20\% (default), 40\%, 60\%, 80\% \\
    Number of WiFi hotspots & 2, 4 (default), 8 \\
        \hline
        \end{tabular}
        }
\vspace{-0.05in}
\end{table}

\begin{table}[b]
\vspace{-0.15in}
\caption{Energy consumption for 3G and WiFi, \cite{Ris++11}.}
    \label{tab:energy}
\vspace{-0.1in}
    \centering
 {\footnotesize
        \begin{tabular}{|c|c|c|}
        \hline
        Technology  &  Transfer (Joule/MB) & Idle (Watt)\\
        \hline \hline
   3G      & 100 & 0 \\
   WiFi    & 5 & 0.77 \\
        \hline
        \end{tabular}
        }
\vspace{-0.1in}
\end{table}

Estimation of the energy consumption uses Table~\ref{tab:energy}, which was obtained from \cite{Ris++11}. We assume that the WiFi interface is activated 20 seconds prior to connecting to the WiFi hotspot.

The evaluation results presented in  this section are based on numerically computing the data transferred over the mobile and WiFi networks for the parameters  in Table~\ref{tab:route_segments_4} and \ref{tab:values}.
The graphs presented show averages and  95\% confidence intervals from 120 runs of each scenario.
Also,  the values in Table~\ref{tab:values} depicted as default are the values of the parameters that do not change in the specific evaluation scenario (graph).


\subsection{Delay tolerant traffic}

In this subsection we discuss results for delay tolerant traffic, where a data object needs to be transferred until the  end of the vehicle's route in Figure~\ref{fig:path}. We compare the following three cases: the procedure that exploits mobility prediction and prefetching (Algorithm~\ref{alg:delaytolerant}), the procedure that exploits only mobility prediction without prefetching, and the case when  prediction is not utilized and the  maximum available mobile throughput  is always used.
The metrics we consider are the percentage of traffic that is offloaded to the mobile network and the energy consumption.

\begin{figure}[tb]
\begin{center}
\begin{tabular}{c}

\begin{minipage}[b]{0.5\linewidth}
\centering
\hspace{-0.22in}
\includegraphics[width=1.7in] {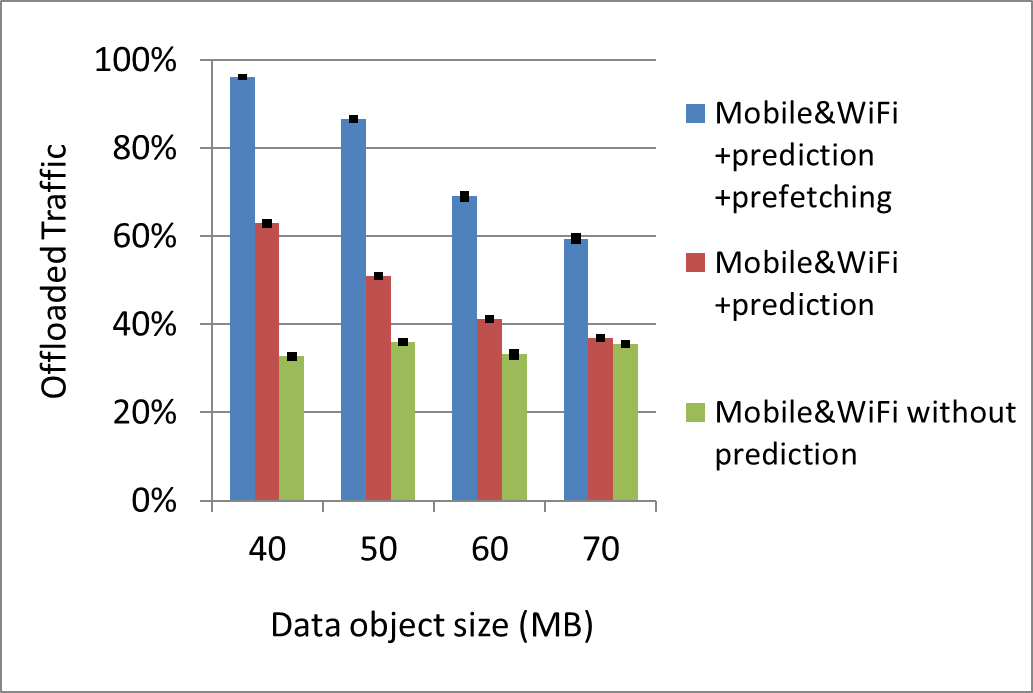}\\
{\footnotesize \hspace{-0.22in} \small{(a) Offloaded traffic}}
\end{minipage}
\begin{minipage}[b]{0.5\linewidth}
\centering
\hspace{-0.22in}
\includegraphics[width=1.7in]{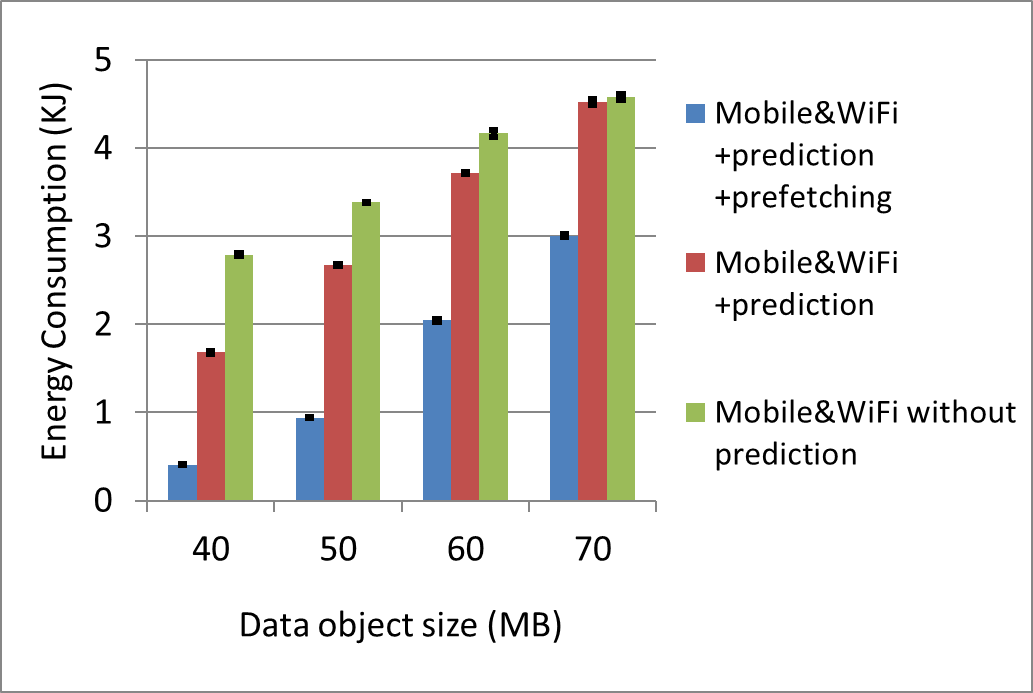}\\
{\footnotesize \hspace{-0.22in} \small{(b) Energy consumption}}
\end{minipage}\

\end{tabular}
\end{center}
\vspace{-.1 in}
\caption[]{\protect \small{Percentage of offloaded traffic and energy consumption as a function of data object size. Delay tolerant traffic.
}}
\label{fig:dt_size}
\vspace{-0.1in}
\end{figure}

\medskip
\noindent
\emph{Data object size:} Figure~\ref{fig:dt_size}(a) shows the percentage of offloaded traffic for different data object sizes. For all data sizes the percentage of offloaded traffic with the prediction + prefetching scheme is more than 65\% higher compared to the case where  prediction and  prefetching is not used. Moreover, the gains are higher for smaller data sizes.
For large data sizes, the performance of the prediction scheme is close to the performance when prediction is not used; this occurs because for large object sizes the mobile network is used close to its maximum throughput, hence prediction is not beneficial.

Figure~\ref{fig:dt_size}(b) shows that the energy efficiency gains  reflect the gains in terms of offloaded traffic. Specifically, the energy efficiency gains with the prediction + prefetching scheme is approximately 85\% for a 40~MB data object size and 35\% for a 70~MB object size. For large data sizes, the energy consumption of the prediction scheme is close to the energy consumption when prediction is not used.

\begin{figure}[tb]
\begin{center}
\begin{tabular}{c}

\begin{minipage}[b]{0.5\linewidth}
\centering
\hspace{-0.22in}
\includegraphics[width=1.7in] {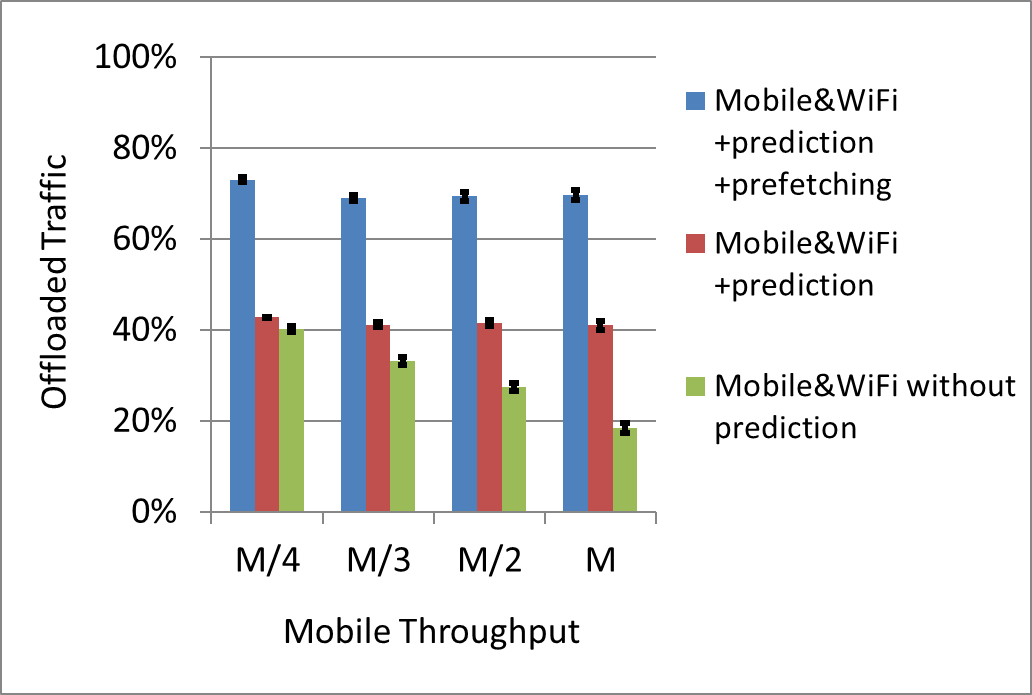}\\
{\footnotesize \hspace{-0.22in}\small{(a) Mobile throughput}}
\end{minipage}
\begin{minipage}[b]{0.5\linewidth}
\centering
\hspace{-0.22in}
\includegraphics[width=1.7in]{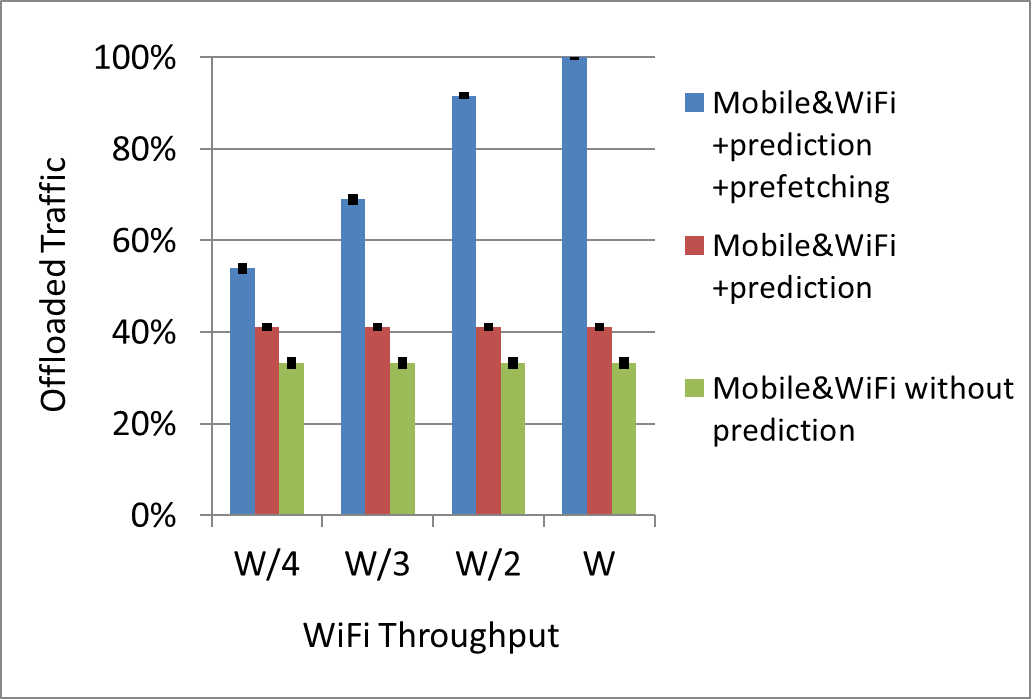}\\
{\footnotesize  \hspace{-0.22in}\small{(b) WiFi throughput}}
\end{minipage} \\ $\,$ \\
\begin{minipage}[b]{0.5\linewidth}
\centering
\hspace{-0.22in}
\includegraphics[width=1.7in]{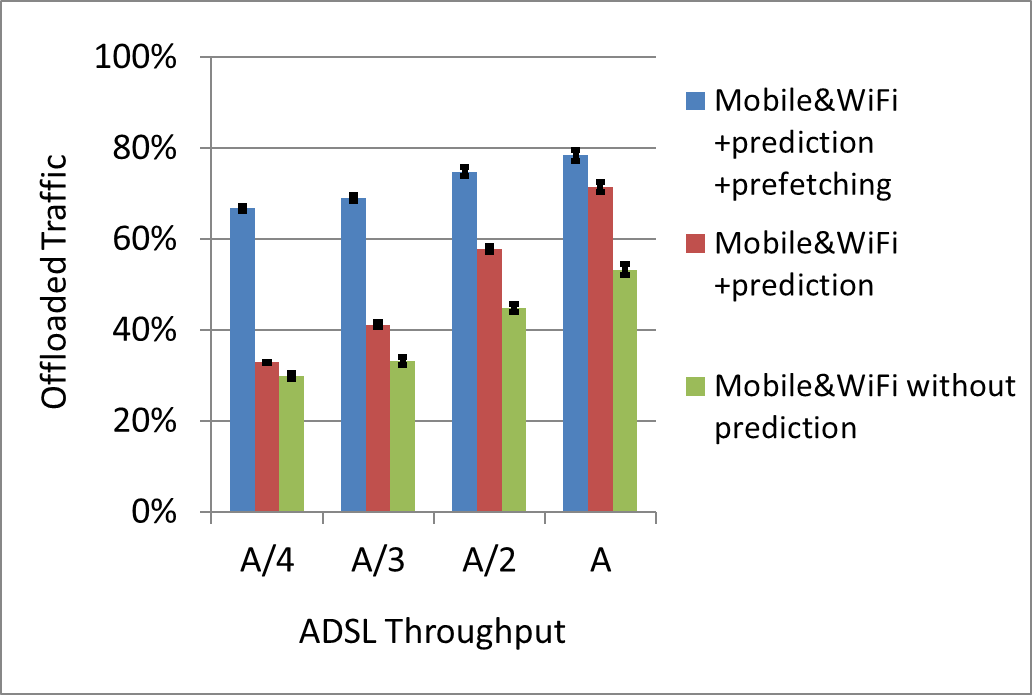}\\
{\footnotesize  \hspace{-0.22in}\small{(c) ADSL throughput}}
\end{minipage}
\begin{minipage}[b]{0.5\linewidth}
\centering
\hspace{-0.22in}
\includegraphics[width=1.7in]{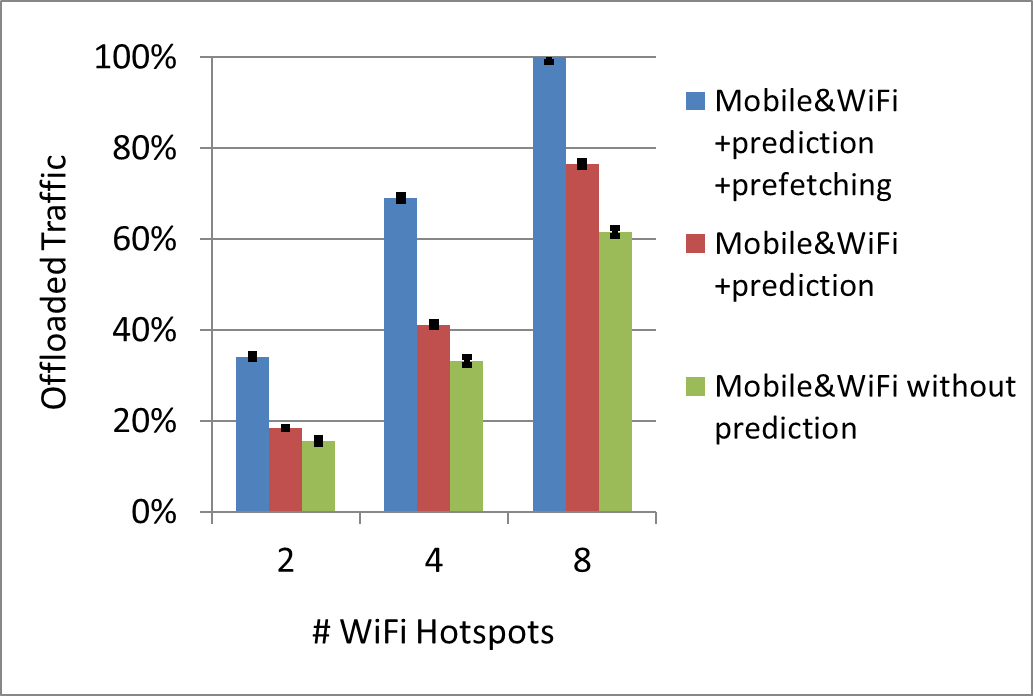}\\
{\footnotesize  \hspace{-0.22in}\small{(d) \# of WiFi hotspots}}
\end{minipage}
\end{tabular}
\end{center}
\vspace{-.1 in}
\caption[]{\protect \small{Percentage of offloaded traffic as a function of mobile, WiFi, and ADSL throughput, and number of WiFi hotspots. Delay tolerant traffic.
}}
\label{fig:dt_thr}
\vspace{-0.1in}
\end{figure}

\medskip
\noindent
\emph{Mobile, WiFi, ADSL backhaul throughput:} Figure~\ref{fig:dt_thr}(a) shows the percentage of offloaded traffic for different mobile throughputs.
As expected, the percentage of offloaded traffic for the prediction + prefetching and the prediction schemes does not depend on the mobile throughput, since these schemes use less than the maximum available mobile throughput. On the other hand, when prediction and prefetching is not used, the percentage of offloaded traffic decreases when the mobile throughput increases.

Figure~\ref{fig:dt_thr}(b) shows the percentage of offloaded traffic for different WiFi throughputs. For the prediction + prefetching scheme the amount of offloaded traffic increases significantly when the WiFi throughput increases. On the other hand, it does not affect the prediction scheme and the case when prediction and prefetching is not used, since in these cases  the amount of offloaded traffic is constrained by the ADSL throughput.

Figure~\ref{fig:dt_thr}(c) shows the  offloaded traffic for different ADSL backhaul throughputs.
When the throughput is low, the performance when only  prediction is used is close to the performance when  prediction is not used; this happens because when the backhaul throughput is low, the mobile network needs to be used more, hence the mobile throughput  is close to its maximum.  On the other hand, when the backhaul throughput is high, then the performance of prediction and prefetching is close to the performance when only prediction is used; this occurs because when the backhaul throughput is high and close to the WiFi throughput, there are smaller gains from prefetching and downloading data from a local cache.

\medskip
\noindent
\emph{Number of WiFi hotspots:} Figure~\ref{fig:dt_thr}(d) shows that for two hotspots, the prediction scheme has similar performance when no prediction is used; this occurs because for few hotspots, the prediction scheme uses mobile throughput close to the maximum. On the other hand, the prediction + prefetching scheme achieves performance which is more than 30\% higher than the prediction scheme and more than 60\% higher than when prediction and prefetching is not used.

\begin{figure}[tb]
\begin{center}
\begin{tabular}{c}

\begin{minipage}[b]{0.5\linewidth}
\centering
\hspace{-0.22in}
\includegraphics[width=1.7in] {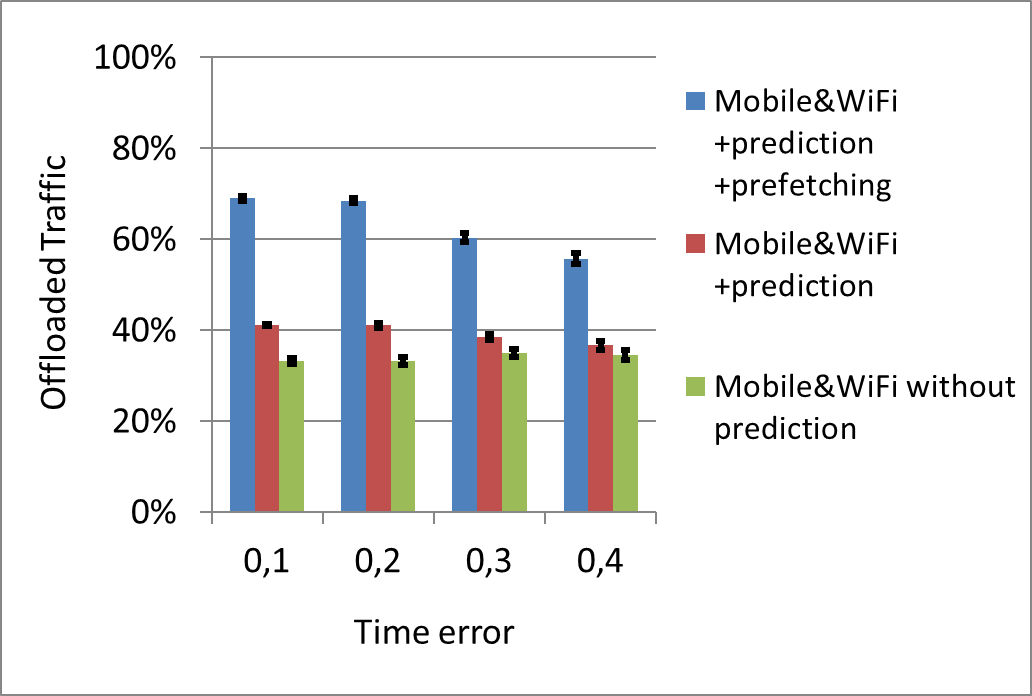}\\
{\footnotesize \hspace{-0.22in}\small{(a) Time error}}
\end{minipage}
\begin{minipage}[b]{0.5\linewidth}
\centering
\hspace{-0.22in}
\includegraphics[width=1.7in]{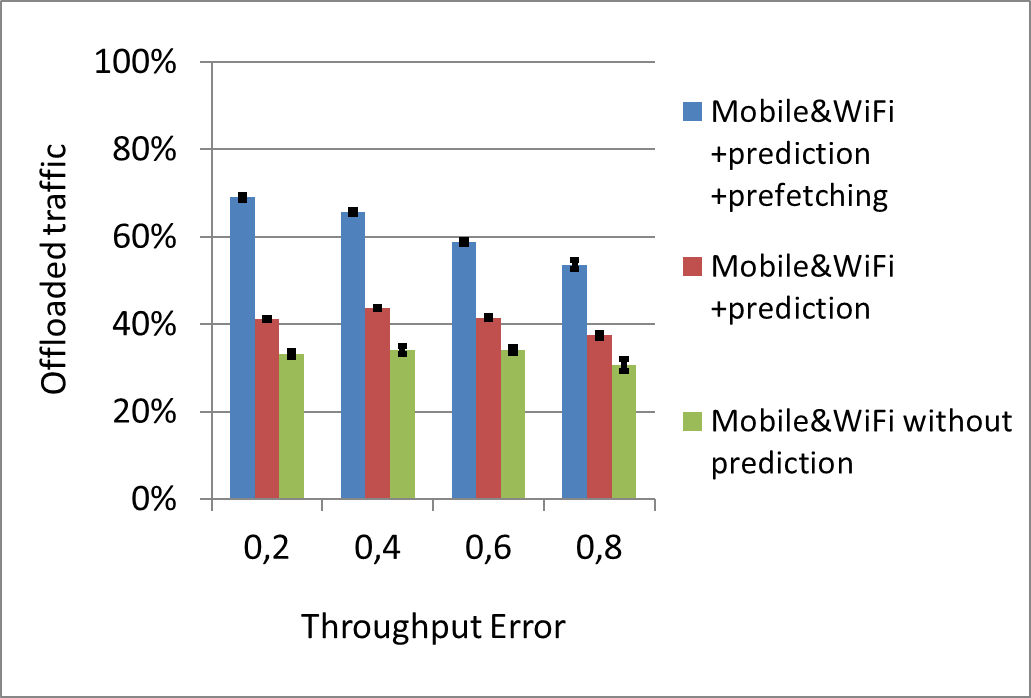}\\
{\footnotesize  \hspace{-0.22in}\small{(b) Throughput error}}
\end{minipage}\
\end{tabular}
\end{center}
\vspace{-.1 in}
\caption[]{\protect \small{Percentage of offloaded traffic as a function of time and throughput error. Delay tolerant traffic.
}}
\label{fig:dt_error}
\vspace{-0.05in}
\end{figure}

\begin{figure}[tb]
\begin{center}
\begin{tabular}{c}

\begin{minipage}[b]{0.5\linewidth}
\centering
\hspace{-0.22in}
\includegraphics[width=1.7in] {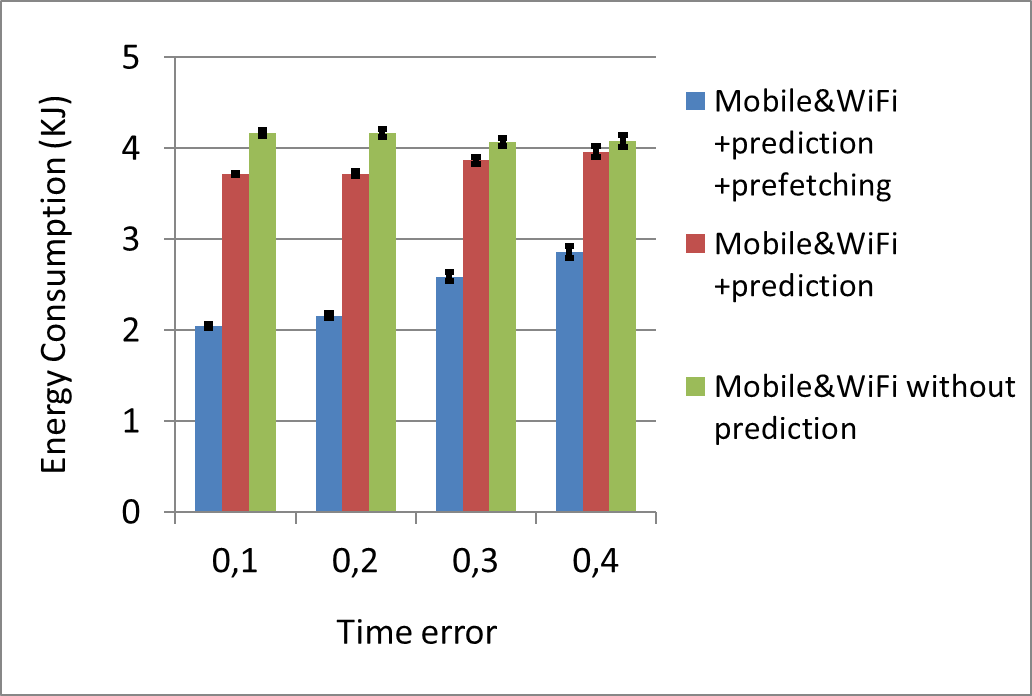}\\
{\footnotesize \hspace{-0.22in}\small{(a) Time error}}
\end{minipage}
\begin{minipage}[b]{0.5\linewidth}
\centering
\hspace{-0.22in}
\includegraphics[width=1.7in]{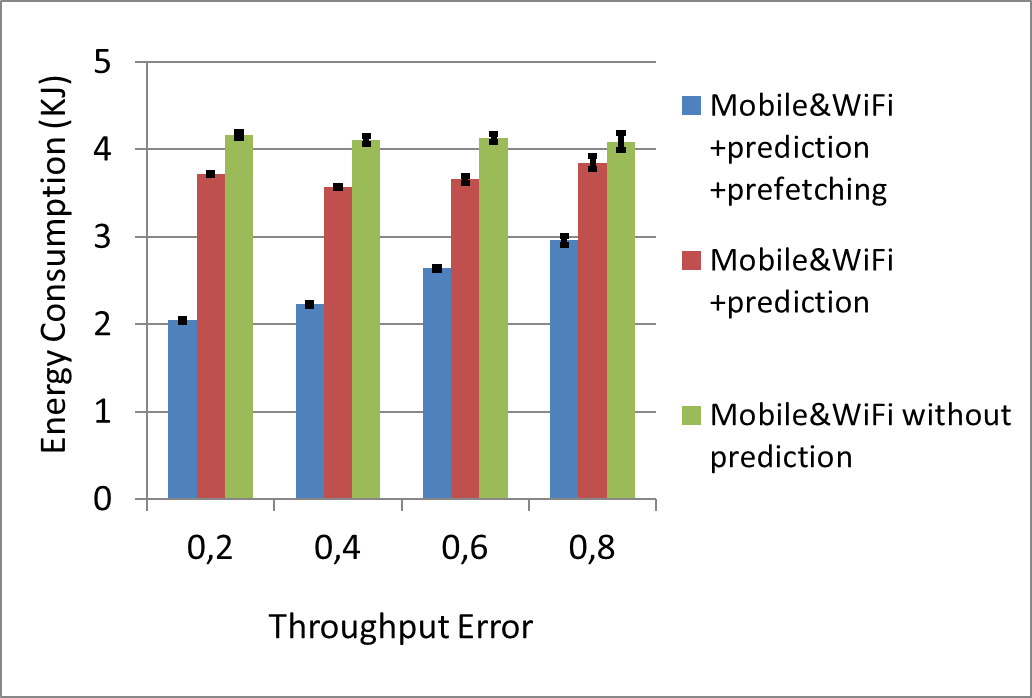}\\
{\footnotesize  \hspace{-0.22in}\small{(b) Throughput error}}
\end{minipage}\

\end{tabular}
\end{center}
\vspace{-.1 in}
\caption[]{\protect \small{Energy consumption as a function of time and throughput error. Delay tolerant traffic.
}}
\label{fig:dt_error_energy}
\vspace{-0.05in}
\end{figure}

\medskip
\noindent
\emph{Time error:} Figure~\ref{fig:dt_error}(a) shows how the percentage of offloaded traffic is affected by the time error. Observe that the performance when prediction and prefetching are used and when only prediction is  used decreases as the  time error increases; this occurs because the time error reduces the effectiveness of prediction and prefetching.
Nevertheless, the offloading percentage when prediction and prefetching are used is more than 60\% higher than the offloading percentage  when  prediction and prefetching are not used, and more than 50\% higher than the offloading percentage when only prediction is used. Figure~\ref{fig:dt_error_energy}(a) shows that the gains in terms of reduced energy consumption follow the gains of the increased amount of offloaded traffic, Figure~\ref{fig:dt_error}(a).

\medskip
\noindent
\emph{Throughput error:} Figure~\ref{fig:dt_error}(b) shows that the throughput error affects the performance of the prediction and prefetching scheme most. Nevertheless, its performance remains more than 40\% higher than  the prediction-only scheme and more than 70\% higher than the case where prediction and prefetching are not used, even when the throughput error is as high as 80\%.
Figure~\ref{fig:dt_error_energy}(b) shows that the gains in terms of reduced energy consumption follow the gains of the increased amount of offloaded traffic, Figure~\ref{fig:dt_error}(b).


\subsection{Delay sensitive traffic}

A key difference compared to delay tolerant traffic is that now the maximum mobile throughput  is always used.
We compare three cases: the procedure that exploits both mobility prediction and prefetching,   the case where prediction and prefetching are not used, and the case where only the mobile network is used.
The performance metric is the delay for transferring a data object and the energy consumption.

\begin{figure}[tb]
\begin{center}
\begin{tabular}{c}

\begin{minipage}[b]{0.5\linewidth}
\centering
\hspace{-0.22in}
\includegraphics[width=1.7in] {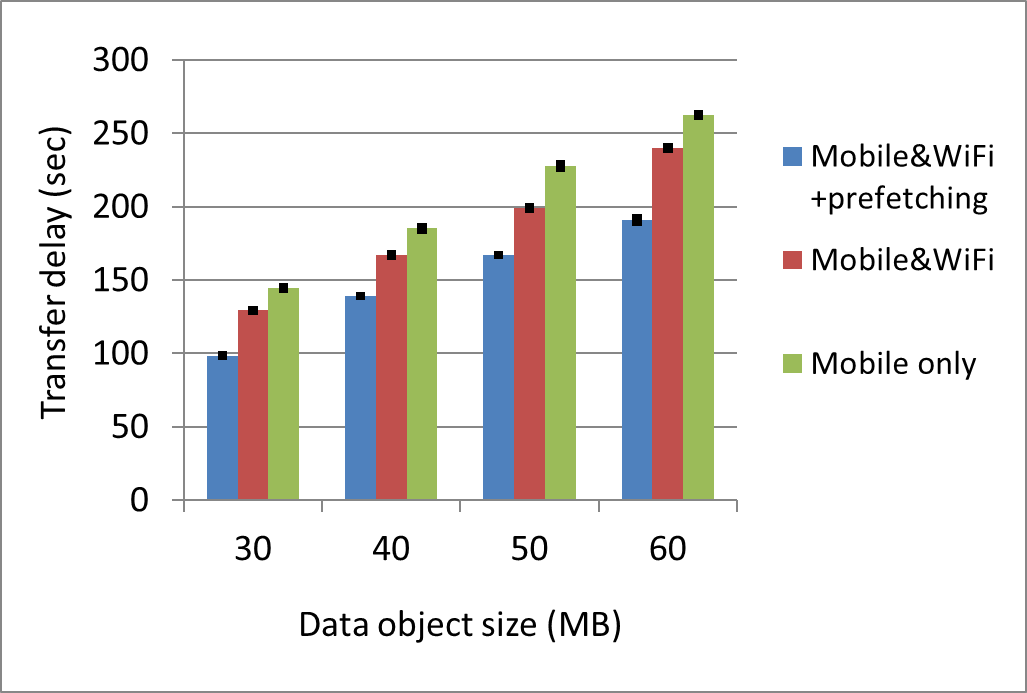}\\
{\footnotesize \hspace{-0.22in}\small{(a) Transfer delay}}
\end{minipage}
\begin{minipage}[b]{0.5\linewidth}
\centering
\hspace{-0.22in}
\includegraphics[width=1.7in]{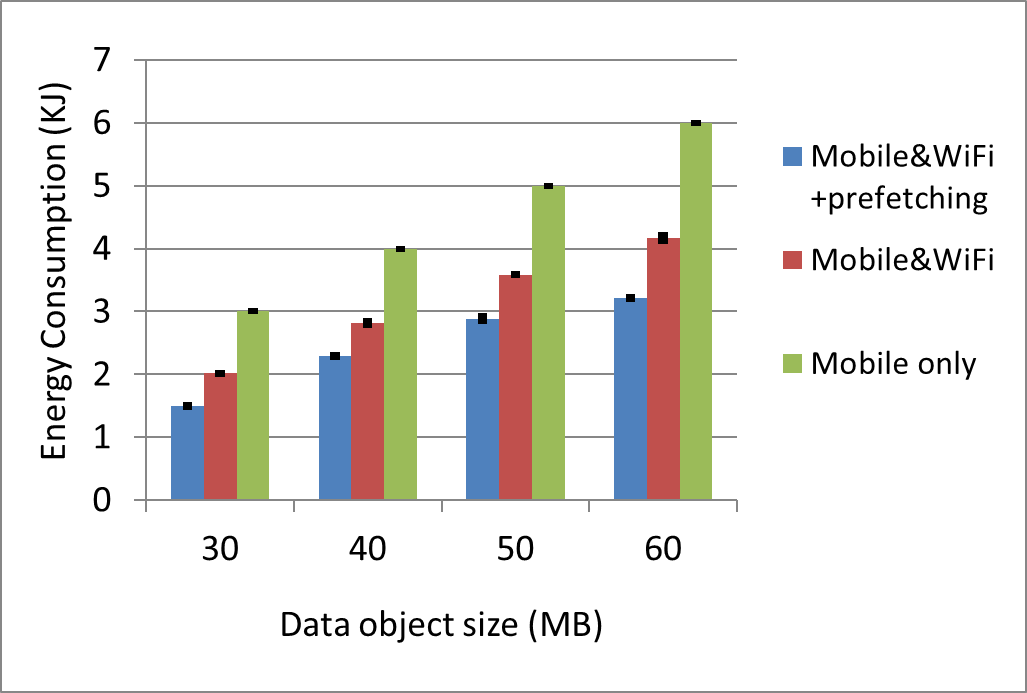}\\
{\footnotesize  \hspace{-0.22in}\small{(b) Energy consumption}}
\end{minipage}\

\end{tabular}
\end{center}
\vspace{-.1 in}
\caption[]{\protect \small{Transfer and energy consumption as a function of data object size. Delay sensitive traffic.
}}
\label{fig:ds_size}
\vspace{-0.1in}
\end{figure}

\medskip
\noindent
\emph{Data object size:} Figure~\ref{fig:ds_size}(a) shows the transfer delay as a function of data object size. Prediction and prefetching achieve a delay that is lower by 25-35\% compared to the case where only the mobile network is used, and 15-25\% compared to  WiFi offloading without prediction and  prefetching.

The energy efficiency gains with prediction and prefetching are approximately 20-25\% compared to the case of WiFi offloading without prediction and prefetching.
The energy gains with prediction and prefetching are even higher compared to when only the mobile network is used: 40-50\%.

\begin{figure}[tb]
\begin{center}
\begin{tabular}{c}

\begin{minipage}[b]{0.5\linewidth}
\centering
\hspace{-0.22in}
\includegraphics[width=1.7in] {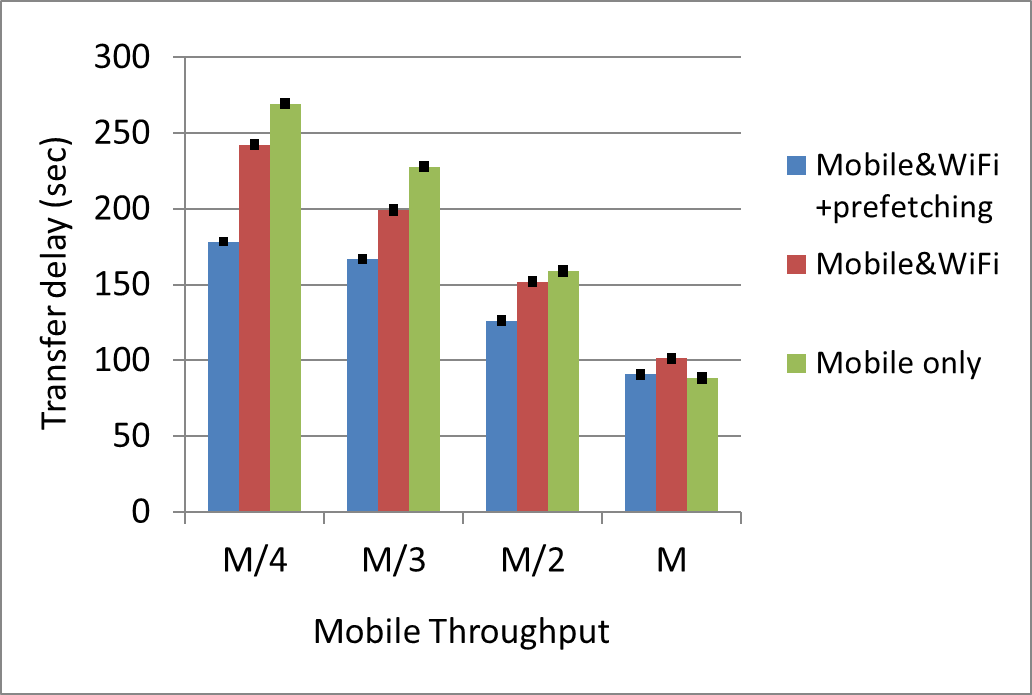}\\
{\footnotesize \hspace{-0.22in}\small{(a) Mobile throughput}}
\end{minipage}
\begin{minipage}[b]{0.5\linewidth}
\centering
\hspace{-0.22in}
\includegraphics[width=1.7in]{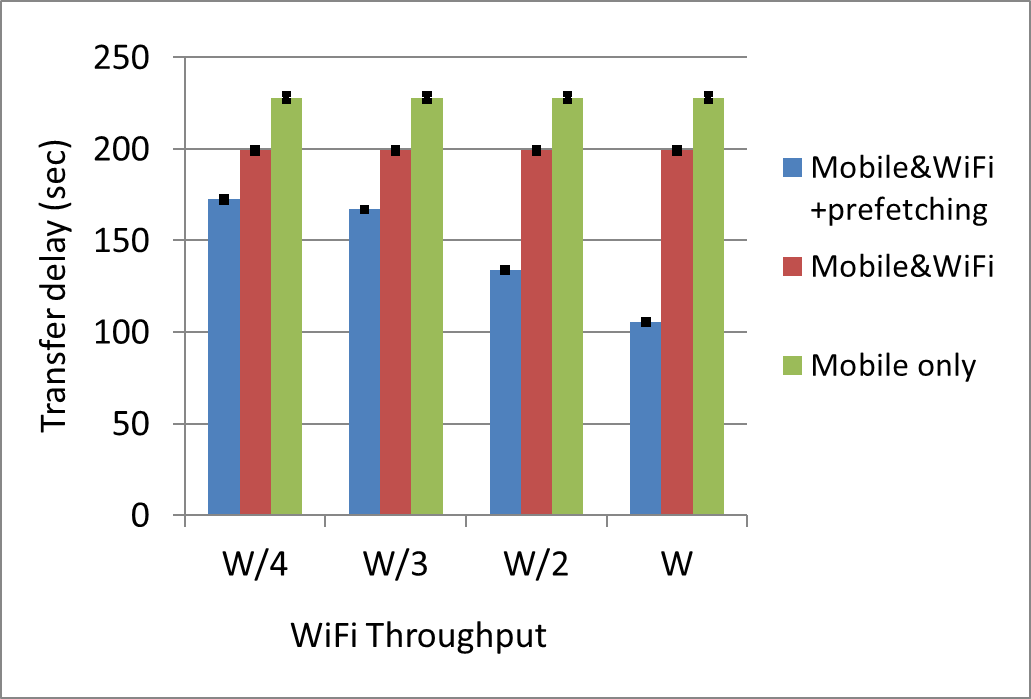}\\
{\footnotesize  \hspace{-0.22in}\small{(b) WiFi throughput}}
\end{minipage} \\ $\,$ \\
\begin{minipage}[b]{0.5\linewidth}
\centering
\hspace{-0.22in}
\includegraphics[width=1.7in]{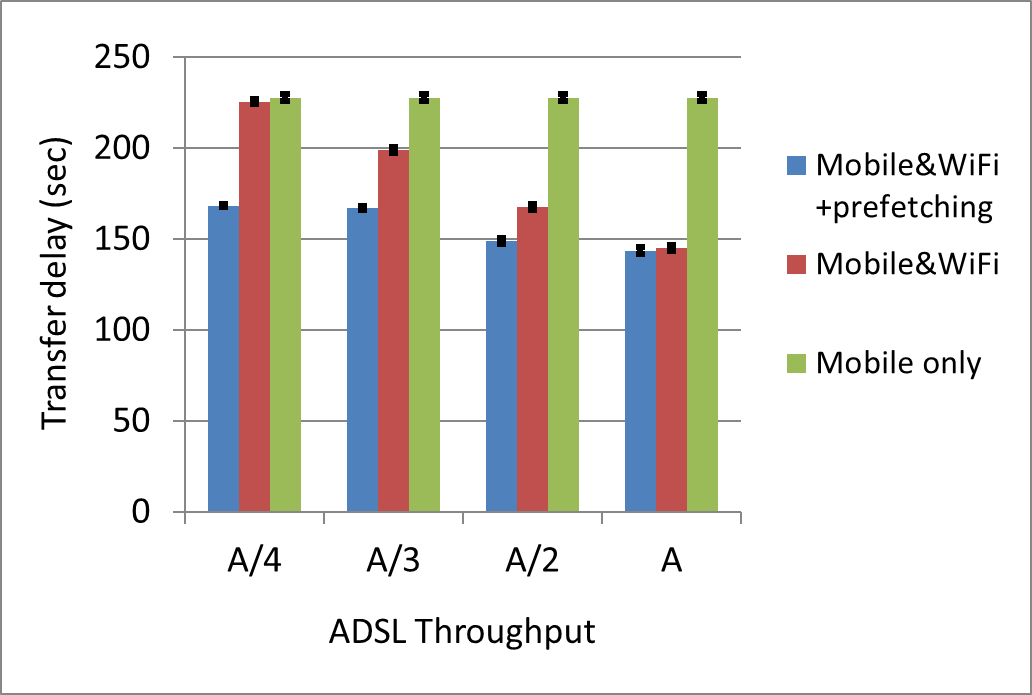}\\
{\footnotesize  \hspace{-0.22in}\small{(c) ADSL throughput}}
\end{minipage}
\begin{minipage}[b]{0.5\linewidth}
\centering
\hspace{-0.22in}
\includegraphics[width=1.7in]{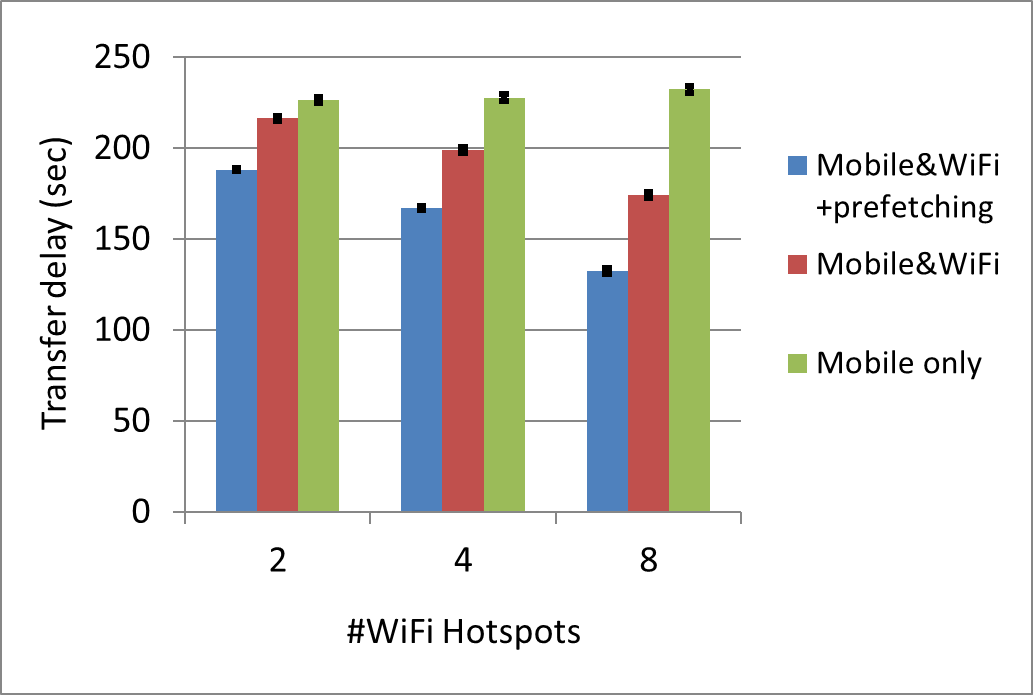}\\
{\footnotesize  \hspace{-0.22in}\small{(c) \# of WiFi hotspots}}
\end{minipage}
\end{tabular}
\end{center}
\vspace{-.1 in}
\caption[]{\protect \small{Transfer delay as a function of mobile, WiFi, and ADSL throughput, and number of WiFi hotspots. Delay sensitive traffic.
}}
\label{fig:ds_thr}
\end{figure}

\medskip
\noindent
\emph{Mobile, WiFi, ADSL backhaul throughput:} Figure~\ref{fig:ds_thr}(a) shows that the transfer delay  gains with prediction and prefetching are higher for a smaller mobile throughput. Moreover, for a high mobile throughput, the transfer delay with offloading can be worse than when only the mobile network is used, when the mobile throughput is higher than the ADSL throughput.

Figure~\ref{fig:ds_thr}(b) shows the transfer delay as a function of the WiFi throughput. As in the case of delay tolerant traffic, the performance of prediction and prefetching in terms of reduced transfer delay increases as the WiFi throughput increases. On the other hand, the transfer delay in the case of WiFi offloading without  prediction and prefetching and when only the mobile network are used is not influenced by the WiFi throughput.

Figure~\ref{fig:ds_thr}(c) shows the influence of the ADSL throughput on the transfer delay. Observe that for a low  ADSL throughput, the performance of WiFi offloading without prediction and prefetching is close to the performance when only the mobile network is used. On the other hand, for high values of the ADSL throughput the performance in the case of  prediction and prefetching is close to the performance in the case of WiFi offloading without prediction and prefetching; this occurs because the gains of prefetching are reduced when the ADSL throughput approaches the WiFi throughput.

\medskip
\noindent
\emph{Number of WiFi hotspots:} Figure~\ref{fig:ds_thr}(d) shows the transfer delay for a different number of hotpots. As expected, the transfer delay improves with prediction and prefetching when the number of hotspots increases: The transfer delay with prediction and prefetching with two hotspots is approximately 13\% and 17\% lower than offloading without prediction and when only the mobile network is used, respectively, while it is approximately 24\% and 43\% lower when there are 8 hotspots.

\begin{figure}[tb]
\begin{center}
\begin{tabular}{c}

\begin{minipage}[b]{0.5\linewidth}
\centering
\hspace{-0.22in}
\includegraphics[width=1.7in] {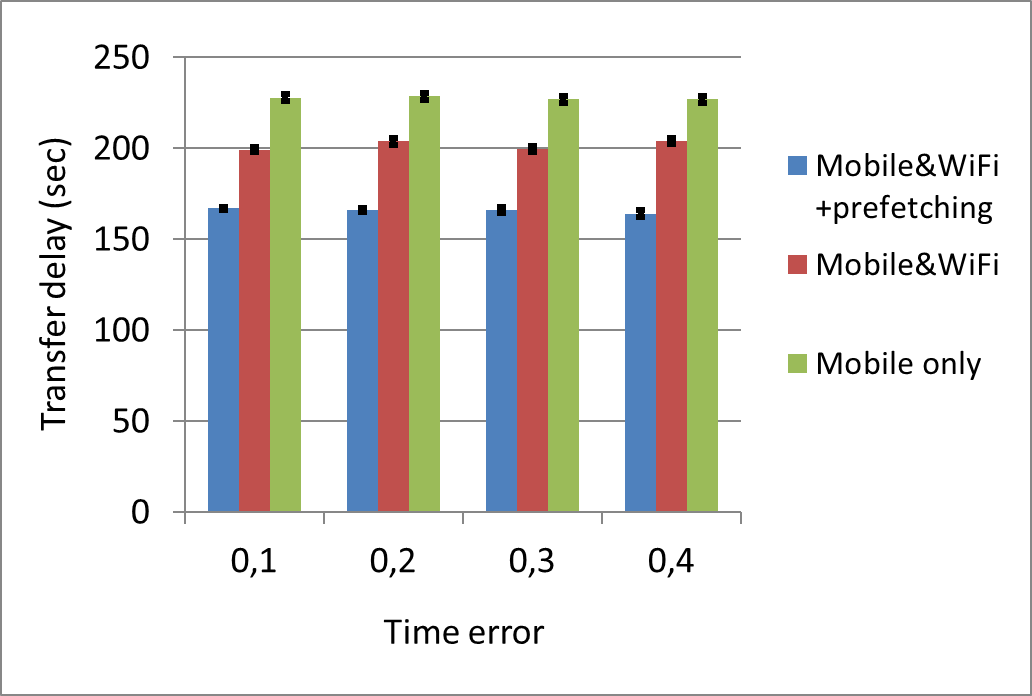}\\
{\footnotesize \hspace{-0.22in}\small{(a) Time error}}
\end{minipage}
\begin{minipage}[b]{0.5\linewidth}
\centering
\hspace{-0.22in}
\includegraphics[width=1.7in]{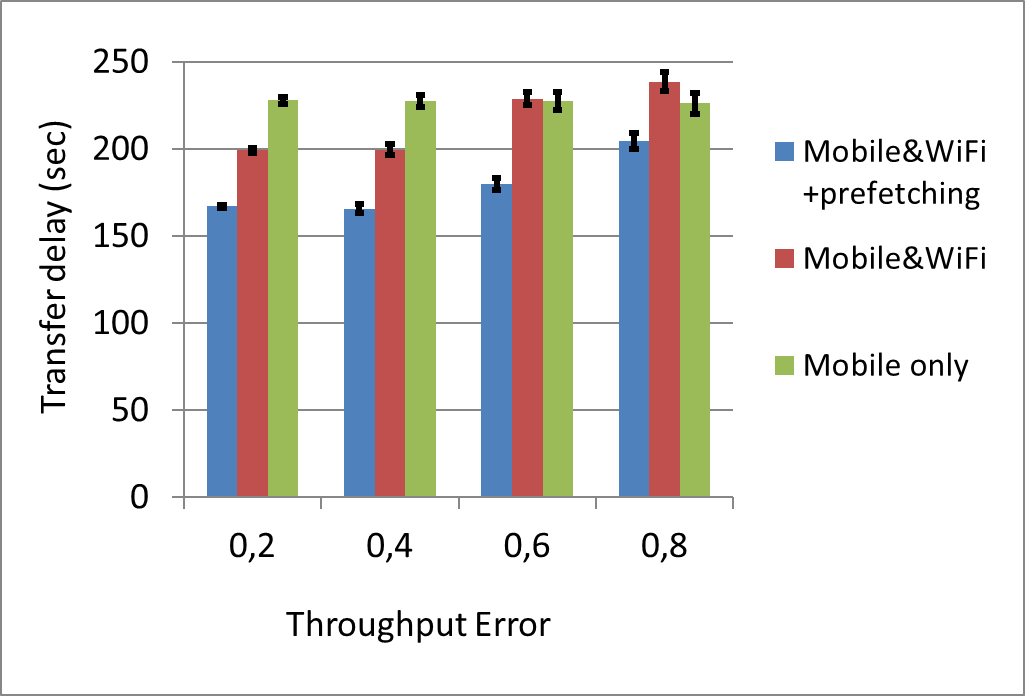}\\
{\footnotesize  \hspace{-0.22in}\small{(b) Throughput error}}
\end{minipage}\

\end{tabular}
\end{center}
\vspace{-.1 in}
\caption[]{\protect \small{Transfer delay as a function of time and throughput error. Delay sensitive traffic.
}}
\label{fig:ds_error}
\vspace{-0.15in}
\end{figure}

\begin{figure}[tb]
\begin{center}
\begin{tabular}{c}

\begin{minipage}[b]{0.5\linewidth}
\centering
\hspace{-0.22in}
\includegraphics[width=1.7in] {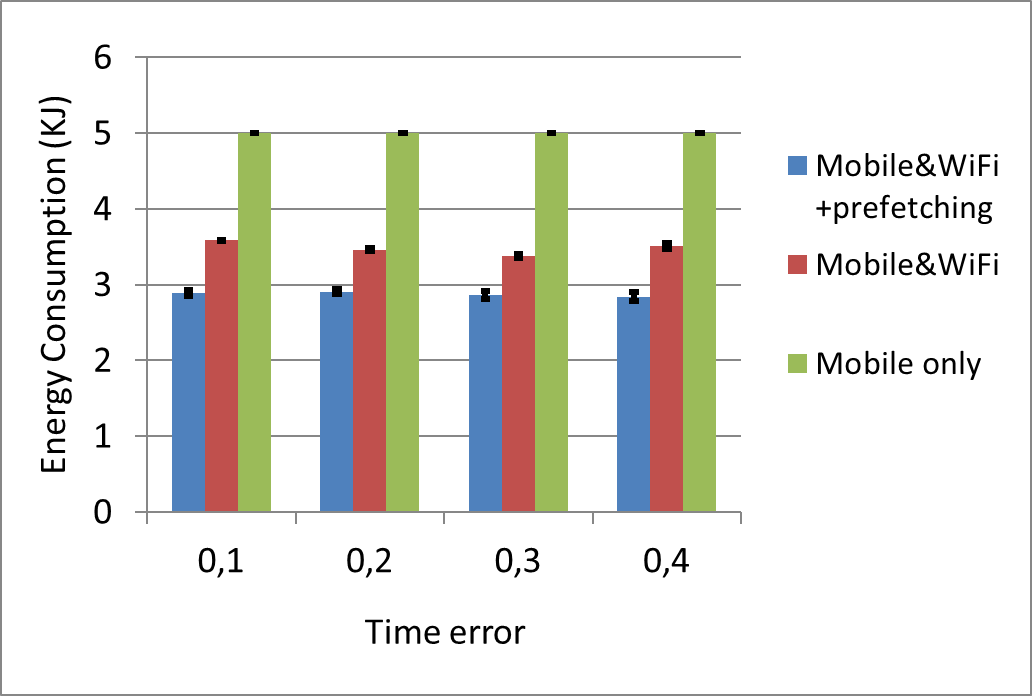}\\
{\footnotesize \hspace{-0.22in}\small{(a) Time error}}
\end{minipage}
\begin{minipage}[b]{0.5\linewidth}
\centering
\hspace{-0.22in}
\includegraphics[width=1.7in]{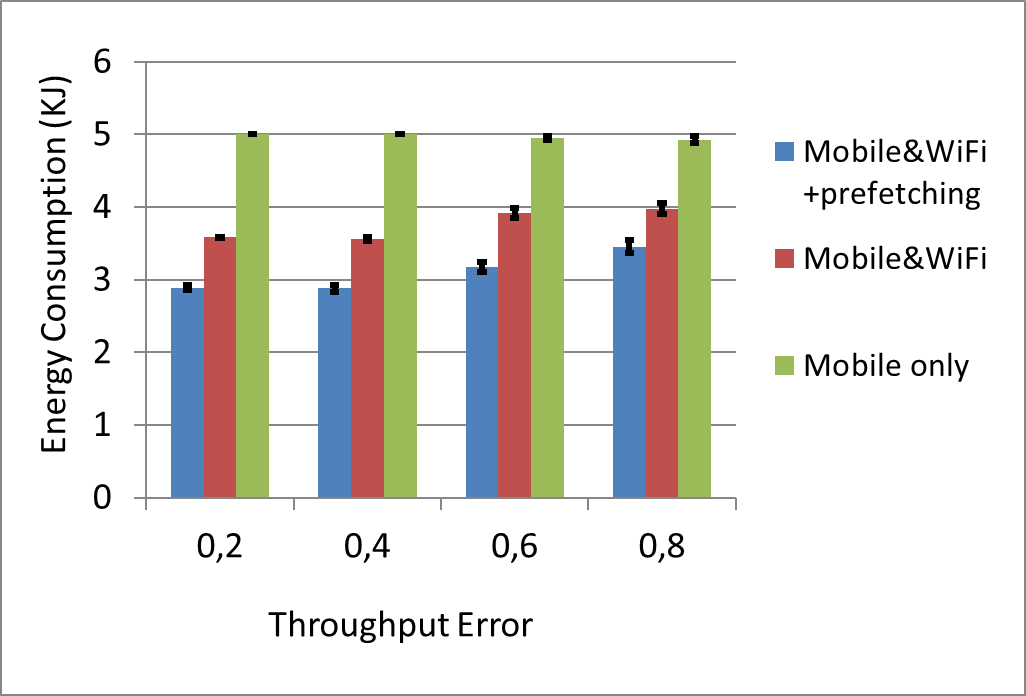}\\
{\footnotesize  \hspace{-0.22in}\small{(b) Throughput error}}
\end{minipage}\

\end{tabular}
\end{center}
\vspace{-.1 in}
\caption[]{\protect \small{Energy consumption as a function of time and throughput error. Delay sensitive traffic.
}}
\label{fig:ds_error_energy}
\vspace{-0.05in}
\end{figure}

\medskip
\noindent
\emph{Time error:} Figure~\ref{fig:ds_error}(a) shows  that as the time error increases, the variability of the transfer delay increases slightly (the 95\% confidence interval is larger), but the average transfer delay for all schemes remains the same. Figure~\ref{fig:ds_error_energy}(a) shows the energy consumption as a function of time errors. Observe that the average energy efficiency gains are independent of the time errors and are relatively higher compared to the transfer delay gains: When prediction and prefetching are used the energy consumption is more than 40\% lower than when only the mobile network is used, whereas the transfer delay reduction  is approximately 27\%.

\medskip
\noindent
\emph{Throughput error:} Figure~\ref{fig:ds_error}(b) shows the influence of the throughput error on the transfer delay. As expected, the transfer delay gains are higher for lower throughput errors; however, observe that some gains still exist with prediction and prefetching even when the throughput error becomes very high (80\%). Also, observe that with a high throughput error the transfer delay when WiFi offloading without prediction and prefetching is used can be higher than when only the mobile network is used.
Figure~\ref{fig:ds_error_energy}(b) shows the energy consumption as a function of throughput errors. Observe that a higher throughput error reduces the energy efficiency gains, which however still remain high: with a 80\% throughput error, prediction and prefetching achieve lower energy consumption by approximately 30\% compared to the case where only the mobile network is used and 13\% when WiFi offloading is used without prediction and prefetching.

\section{Conclusions and Future Work}
\label{sec:conclusions}

We have presented a comprehensive evaluation of procedures that exploit mobility prediction and prefetching to enhance mobile data offloading, for both delay tolerant and delay sensitive traffic. Our evaluation is in terms of the amount of offloaded traffic, the data transfer delay, and the energy consumption, and shows how the performance depends on various factors, such as the data object size, the mobile, WiFi, and ADSL backhaul throughput, the number of WiFi hotspots, and the robustness of the proposed procedures to time and throughput estimation errors.
Future work includes implementing a prototype to demonstrate the gains of the proposed offloading procedures. Moreover, we are extending the procedures to  allow different tradeoffs between the delay, the amount of offloaded traffic, and the energy efficiency, and to exploit prediction and prefetching for streaming video.

\mynotex{
\begin{itemize}
\item streaming video
\item prototype
\end{itemize}
}

\mynotex{
\begin{itemize}
\item evaluate energy gains
\item implement prototype, extending the OptiPath application. In a first phase this will involve only the procedure to exploit mobility prediction, since prefetching needs support (caching) at the WiFi hotspots.
\end{itemize}
}




\bibliographystyle{abbrv}

{
\bibliography{pref} }

\end{document}